\documentclass[aps, pra, showpacs,superscriptaddress,
unsortedaddress, twocolumn, preprintnumbers] {revtex4}
\usepackage{graphicx}
\usepackage{calc}
\usepackage{amsmath}
\usepackage{amssymb}

\newcommand{\Tr}{\mbox{\rm Tr}}

\begin{document}
\preprint{PHYSICAL REVIEW A {\bf 83}, 032119 (2011); 85, 049904(E) (2012)}
\author{A. A. Semenov}
\email[E-mail address: ]{sem@iop.kiev.ua}%

\affiliation{Institut f\"ur Physik, Universit\"{a}t Rostock,
Universit\"{a}tsplatz 3, D-18051 Rostock, Germany}
\affiliation{Institute of Physics, National Academy of Sciences of
Ukraine,
Prospect Nauky 46, UA-03028 Kiev, Ukraine}%
\affiliation{Bogolyubov Institute for Theoretical Physics, National
Academy of Sciences of Ukraine,\\ Vul. Metrologichna 14-b, UA-03680
Kiev, Ukraine}%
\author{W. Vogel}
\affiliation{Institut f\"ur Physik, Universit\"{a}t Rostock,
Universit\"{a}tsplatz 3, D-18051 Rostock, Germany}

\title{Fake violations of the quantum Bell-parameter bound}

\begin{abstract}
Shortcomings of experimental techniques are usually assumed to 
diminish nonclassical properties of quantum systems. Here it is
demonstrated that this standard assumption is not true in general.
It is theoretically shown that the inability to resolve different
photon numbers in photodetection may pseudo-increase a measured Bell
parameter. Under proper conditions one even pseudo-violates the
quantum Cirel'son bound of the Bell parameter, the corresponding density
operator fails to be positive semi-definite. This paradox can be
resolved by appropriate squash models.
\end{abstract}

\pacs{03.65.Ud, 42.50.Xa, 42.65.Lm}

\maketitle

\section{Introduction}
\label{Introduction}

Quantum mechanics is not a local realistic theory; this means that
values of observables may not be predefined. This conclusion from the famous work
by Einstein, Podolsky, and Rosen~\cite{EPR} still attracts a great deal of attention
from physicists. Bell \cite{Bell} has proposed a formal framework for this discussion.
He has formulated inequalities that are valid for the local realism, but they are
violated by quantum physics. According to this, the Bell parameter is less than
the value of $2$ for any local realistic theory. In the quantum world this parameter
may exceed this threshold up to the level of $2\sqrt{2}$, known as the Cirel'son
bound~\cite{Tsirelson}. The increasing interest in this subject has also stimulated a
variety of experiments, see e.g.~\cite{Aspect}.

An intriguing question is whether the Bell parameter may exceed the Cirel'son
quantum bound. Popescu and Rohrlich have considered the consequences of two axioms:
nonlocality and relativistic causality~\cite{Popesku}. They have shown that
as a consequence of the axioms, quantum mechanics appears as a particular
representative of a more general theory. The latter includes the possibility of
violating the Cirel'son quantum bound, with the maximum value of the Bell parameter
being $4$. Another situation has been considered by Cabello~\cite{Cabello}. He has
demonstrated that two qubits of a three-qubit system may also violate the Cirel'son
quantum bound up to the level of $4$ based on the standard quantum
theory.

The shortcomings of experimental techniques, including
losses, noise, etc., play two different roles in Bell-type
experiments. First, they lead to a decrease of the Bell parameter; 
even for the Bell states the violation is no longer maximal (see,
e.g., \cite{Fedrizzi}). Second, small values of the detection
efficiency are the subject of a loophole for local realism in Bell
inequalities~\cite{Pearle}. For a discussion of other loopholes we
refer the reader to~\cite{Kwiat}. The mentioned
disadvantages also result in problems with the implementation of
quantum-key distribution (QKD) protocols (cf., e.g., \cite{QuantumInf}).

A special example of such shortcomings is the
impossibility to resolve between different numbers of photons. For
on-off detectors it is only possible to distinguish between the presence
and absence of detected photons. In addition, some standard experiments
restrict the interpretation to low-dimensional Hilbert spaces, while the real
electromagnetic field is characterized by an infinite-dimensional
Hilbert space. In this case, the measurement procedure and further
postprocessing in fact maps or squashes the higher-dimensional
Hilbert space onto a low-dimensional one. If such a
procedure is consistent, an appropriate squash model exists~\cite{squashing}.

The aim of this paper is to show that the straightforward
measurement procedure in Bell-type experiments may result in a
fake enhancement of nonclassical properties. Moreover, we predict
that even the limits of quantum physics may be pseudoviolated and that the
Bell parameter exceeds the quantum Cirel'son bound. In such cases, the
reconstructed density operator fails to be positive semidefinite. Of course, such
fake effects do not mean that we predict violations of
quantum physics. Nevertheless, the correct postprocessing, which is
consistent with a properly defined squash model, resolves this
problem and yields acceptable results.
A clear understanding of this problem is of importance for the
security of quantum communication.

Our paper is organized as follows. In Sec.~\ref{BellTypeExperiment} we consider the
fake violation of the Cirel'son bound of the Bell parameter by using on-off detectors.
A resolution of this paradoxial result by a proper squash model is given in
Sec.~\ref{DoubleClickEvents}. An alternative way of demonstrating the fake violation
of quantum physics under study by the reconstruction of ill-defined two-qubit density
operators is considered in Sec.~\ref{TwoQubitDensityOperator}. In
Sec.~\ref{SummaryAndConclusions} we give a summary and some conclusions.

\section{Bell-type experiment}
\label{BellTypeExperiment}

\begin{figure}[ht!]
\includegraphics[clip=,width=\linewidth]{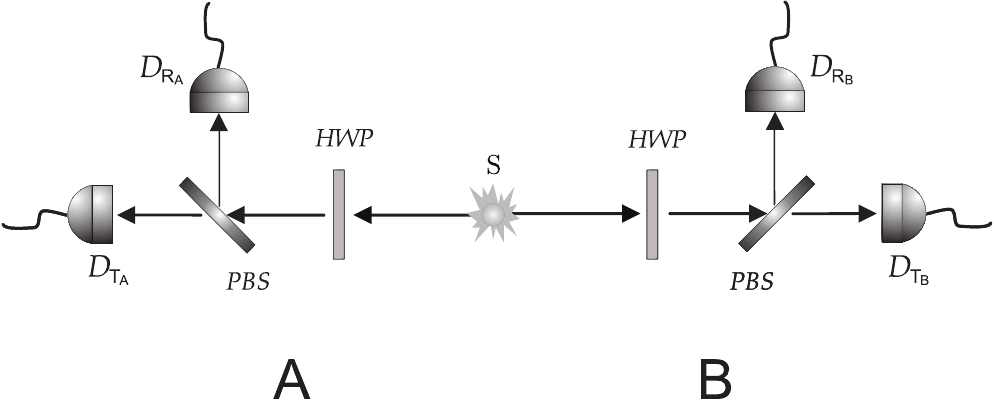}
\caption{\label{Fig1} A typical experimental setup for checking the
violation of Bell inequalities. The source $S$ produces entangled
photon pairs. The polarization analyzers $A$ and $B$ consist of
half-wave plates ($HWP$) polarizing beam splitters ($PBS$), and two
pairs of detectors: $D_\mathrm{T_A}$ ($D_\mathrm{T_B}$) for the transmitted signal
and $D_\mathrm{R_A}$ ($D_\mathrm{R_B}$)  for the reflected signal at the $A$ site
($B$ site).}
\end{figure}

Let us start with a standard experimental setup briefly sketched in
Fig.~\ref{Fig1}. The source of entangled photons irradiates into
four modes: the horizontal and vertical polarized modes at site
$A$ and similar modes at site $B$. Typically each polarization
analyzer consists of a half-wave plate, which can change the
polarization direction to the angles $\theta_\mathrm{A}$ and
$\theta_\mathrm{B}$ at $A$ and $B$ sites, respectively, a
polarization beam splitter, and two detectors for the reflected and
transmitted modes.

Let $P_\mathrm{i_\mathrm{A}, i_\mathrm{B}}\left(\theta_\mathrm{A},
\theta_\mathrm{B}\right)$ be the probability to get a click in one
of the detectors $i_\mathrm{A}=\{\mathrm{T_A},\mathrm{R_A}\}$ at 
site $A$ and in one of the detectors
$i_\mathrm{B}=\{\mathrm{T_B},\mathrm{R_B}\}$ at site $B$ for the
angles of the polarization analyzers $\theta_\mathrm{A}$ and
$\theta_\mathrm{B}$, respectively. The correlation coefficient is
given by
\begin{equation}
E\left(\theta_\mathrm{A}, \theta_\mathrm{B}\right) =
\frac{P_\mathrm{same}\left(\theta_\mathrm{A},
\theta_\mathrm{B}\right)-P_\mathrm{different}\left(\theta_\mathrm{A},
\theta_\mathrm{B}\right)}{P_\mathrm{same}\left(\theta_\mathrm{A},
\theta_\mathrm{B}\right)+P_\mathrm{different}\left(\theta_\mathrm{A},
\theta_\mathrm{B}\right)},\label{correlation}
\end{equation}
where
\begin{equation}
P_\mathrm{same}\left(\theta_\mathrm{A},
\theta_\mathrm{B}\right)=P_\mathrm{\mathrm{T_A},
\mathrm{T_B}}\left(\theta_\mathrm{A},
\theta_\mathrm{B}\right)+P_\mathrm{\mathrm{R_A},
\mathrm{R_B}}\left(\theta_\mathrm{A}, \theta_\mathrm{B}\right),
\end{equation}
is the probability to get clicks on both detectors in the
transmission channels or both detectors in the reflection channels,
and
\begin{equation}
P_\mathrm{different}\left(\theta_\mathrm{A},
\theta_\mathrm{B}\right)=P_\mathrm{\mathrm{T_A},
\mathrm{R_B}}\left(\theta_\mathrm{A},
\theta_\mathrm{B}\right)+P_\mathrm{\mathrm{R_A},
\mathrm{T_B}}\left(\theta_\mathrm{A}, \theta_\mathrm{B}\right)
\label{pdiff}
\end{equation}
is the probability to get clicks on the detectors in the
transmission channel at one site and the reflection channel at
another site. The Clauser-Horne-Shimony-Holt (CHSH) Bell-type
inequality \cite{CHSH} states that the parameter%
\setlength\arraycolsep{0pt}
\begin{eqnarray}
\mathcal{B}&=&\left|E\left(\theta_\mathrm{A}^{(1)},
\theta_\mathrm{B}^{(1)}\right)-E\left(\theta_\mathrm{A}^{(1)},
\theta_\mathrm{B}^{(2)}\right)\right|\label{BellParameter}
\\&+&\left|E\left(\theta_\mathrm{A}^{(2)},
\theta_\mathrm{B}^{(2)}\right)+E\left(\theta_\mathrm{A}^{(2)},
\theta_\mathrm{B}^{(1)}\right)\right|\nonumber,
\end{eqnarray}
also referred to as the Bell parameter, cannot exceed the value of $2$
for local-realistic theories. In quantum theory it is bounded by the
Cirel'son bound of
$2\sqrt{2}$. As shown below, this bound may be violated in the
presence of experimental imperfections.

According to the photodetection theory \cite{PhotoDetection}, the
probability $P_\mathrm{i_\mathrm{A},
i_\mathrm{B}}\left(\theta_\mathrm{A}, \theta_\mathrm{B}\right)$ for
on-off detectors is given by
\begin{equation}
P_\mathrm{i_\mathrm{A}, i_\mathrm{B}}\left(\theta_\mathrm{A},
\theta_\mathrm{B}\right)=\sum\limits_{n,m=1}^{+\infty}\Tr\left(
\hat{\Pi}_{i_\mathrm{A}}^{(n)}\hat{\Pi}_{i_\mathrm{B}}^{(m)}
\hat{\Pi}_{j_\mathrm{A}}^{(0)}\hat{\Pi}_{j_\mathrm{B}}^{(0)}\hat\varrho\right),
\label{Probability}
\end{equation}
$i_\mathrm{A}\neq j_\mathrm{A}$, $i_\mathrm{B}\neq j_\mathrm{B}$,
where $\hat\varrho$ is the density operator of the light at the
input ports of the polarization analyzers and
\begin{equation}
\hat{\Pi}_{i_\mathrm{A(B)}}^{(n)}=:\frac{\left(\eta\,
\hat{n}_{i_\mathrm{A(B)}}\,+N_\mathrm{nc}\right)^n}{n!}\,
\exp\left(-\eta\,\hat{n}_{i_\mathrm{A(B)}}-N_\mathrm{nc}\right):\label{POVM}
\end{equation}
is the positive operator-valued measure in the presence of non-unit
efficiency $\eta$ and the mean number of noise counts
$N_\mathrm{nc}$ (originated from the internal dark counts and the
background radiation), $::$ means normal ordering; see \cite{Semenov}. 
For simplicity we assume the detection efficiencies and noise-count rates to be equal
for all detectors. The photon-number operator in the channel
$i_\mathrm{A(B)}$, $\hat{n}_{i_\mathrm{A(B)}}=\hat{a}^\dag_{i_\mathrm{A(B)}}\hat{a}_{
i_\mathrm{A(B)}}$ can be expressed by the horizontal and vertical modes,
$\hat{a}_{\scriptscriptstyle \mathrm{H_{A(B)}}}$ and
$\hat{a}_{\scriptscriptstyle \mathrm{V_{A(B)}}}$, respectively,
using the input-output relations for the polarization analyzers:
\begin{eqnarray}
\hat{a}_{\scriptscriptstyle
T_\mathrm{A(B)}}=\hat{a}_{\scriptscriptstyle
\mathrm{H_{A(B)}}}\cos\theta_\mathrm{A(B)}+
\hat{a}_{\scriptscriptstyle \mathrm{V_{A(B)}}}\sin\theta_\mathrm{A(B)}\label{IOR1},\\
\hat{a}_{\scriptscriptstyle
R_\mathrm{A(B)}}=-\hat{a}_{\scriptscriptstyle
\mathrm{H_{A(B)}}}\sin\theta_\mathrm{A(B)}+
\hat{a}_{\scriptscriptstyle
\mathrm{V_{A(B)}}}\cos\theta_\mathrm{A(B)}\label{IOR2}.
\end{eqnarray}

In the case when the source $S$ generates the perfect Bell state,
the Bell parameter $\mathcal{B}$, calculated by using
Eqs.~(\ref{correlation})-(\ref{IOR2}), cannot exceed the Cirel'son
bound of $2\sqrt{2}$. However, realistic sources produce, as a rule, more
complicated states. Consider a parametric down-conversion (PDC)
process of entangled-photon generation~\cite{Kwiat, Ma, Kok,
Popescu}. Such a source, for example, has been recently used for
transferring entanglement over a long-distance free-space
channel~\cite{Fedrizzi}. The state
$\hat\varrho=\left|\Psi\right\rangle\left\langle\Psi\right|$ ,
emitted by the PDC source, is of the form~\cite{Ma, Kok}
\begin{equation}
\left|\Psi\right\rangle=(\cosh\chi)^{-2}\sum\limits_{n=0}^{+\infty}
\sqrt{n+1}\tanh^n\chi\left|\Phi_n\right\rangle,\label{PDC1}
\end{equation}
where $\chi$ is the squeezing parameter and
\begin{eqnarray}
&&\left|\Phi_n\right\rangle=\label{PDC2}\\
&&\frac{1}{\sqrt{n+1}}\sum\limits_{m=0}^{n}\left(-1\right)^m
\left|n-m\right\rangle_\mathrm{H_A}\left|m\right\rangle_\mathrm{V_A}
\left|m\right\rangle_\mathrm{H_B}\left|n-m\right\rangle_\mathrm{V_B}\nonumber.
\end{eqnarray}
For small $\chi$, one often restricts the consideration to
$\mbox{n=1}$, representing a Bell state. Note that it was already
mentioned in~\cite{Kwiat}, that in order to close loopholes for
local realism one should discriminate between photon numbers while
using such sources.

An analysis of higher-term contributions in~\cite{Kok, Popescu}
consists of the restriction to a few terms in the
series~(\ref{PDC1}). However, since this state is Gaussian, we may
obtain the exact analytical expression for Eq.~(\ref{Probability})
which enables us to provide a strict analysis of the problem, as done in
Ref.~\cite{Semenov2}. For simplicity, we suppose that the losses for
all four modes are equal. It is possible to show that in this case
one can include them into the detection losses and describe all the
losses by the efficiency $\eta$ in Eq.~(\ref{POVM}).

After straightforward algebra (cf.~Ref.~\cite{Semenov2}) by using
the state~(\ref{PDC1}) in Eq.~(\ref{Probability}), the probability
$P_\mathrm{i_\mathrm{A}, i_\mathrm{B}}\left(\theta_\mathrm{A},
\theta_\mathrm{B}\right)$ for the state~(\ref{PDC1}) can be written as
\begin{eqnarray}
P_\mathrm{i_\mathrm{A}, i_\mathrm{B}}\left(\theta_\mathrm{A},
\theta_\mathrm{B}\right)&=&\left(1-\tanh^2\chi\right)^4e^{-4N_\mathrm{nc}}
\label{ProbabilitySpecial}\\
&\times&\left[\frac{e^{2N_\mathrm{nc}}}{C_\mathrm{0}+2C_\mathrm{1}+C_\mathrm{
i_\mathrm { A },i_\mathrm{B}}
}- \frac{2e^{N_\mathrm{nc}}}{C_\mathrm{0}+C_\mathrm{1}}
+\frac{1}{C_\mathrm{0}}\right].\nonumber
\end{eqnarray}
Here
\begin{equation}
C_\mathrm{0}=\left\{\eta^2\tanh^2\chi-
\left[1+\left(\eta-1\right)\tanh^2\chi\right]^2\right\}^2,\label{C0}
\end{equation}
\begin{eqnarray}
C_\mathrm{1}&=&\eta\left(1-\eta\right)\left(1-\tanh^2\chi\right)\tanh^2\chi\label{C1}
\\
&\times&\left\{\eta^2\tanh^2\chi-
\left[1+\left(\eta-1\right)\tanh^2\chi\right]^2\right\},\nonumber
\end{eqnarray}
\begin{eqnarray}
C_\mathrm{T_A,T_B}=C_\mathrm{R_A,R_B}&=&\eta^2\tanh^2\chi\left(1-\tanh^2\chi\right)
^2\label{Csame}\\
&\times&\left[\left(1-\eta\right)^2 \tanh^2\chi-\sin^2
\left(\theta_\mathrm{A}-\theta_\mathrm{B}\right)\right],\nonumber
\end{eqnarray}
\begin{eqnarray}
C_\mathrm{T_A,R_B}=C_\mathrm{R_A,T_B}&=&\eta^2\tanh^2\chi\left(1-\tanh^2\chi\right)
^2\label{Cdifferent}\\
&\times&\left[\left(1-\eta\right)^2 \tanh^2\chi-\cos^2
\left(\theta_\mathrm{A}-\theta_\mathrm{B}\right)\right].\nonumber
\end{eqnarray}
Now we may insert Eq.~(\ref{ProbabilitySpecial}) in
Eq.~(\ref{correlation}) and apply the result in
Eq.~(\ref{BellParameter}) for the analysis of the Bell parameter.

\begin{figure}[ht!]
\includegraphics[clip=,width=\linewidth]{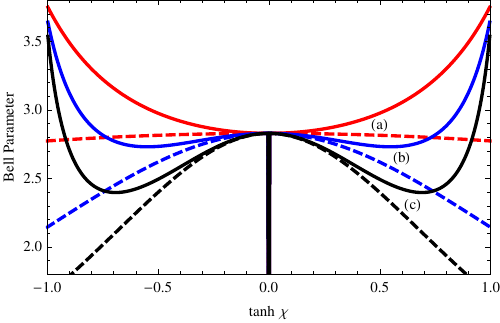}
\caption{\label{Fig2} (Color online) Maximal value of the Bell
parameter $\mathcal{B}$, obtained for on-off (solid lines) and
photon-number-resolving (dashed lines) detectors, vs the parameter
$\tanh\chi$ for the mean value of noise counts
$N_\mathrm{nc}=10^{-6}$ and the efficiencies (a) $\eta=0.9$, (b)
$\eta=0.6$, and (c) $\eta=0.4$. The minimum at $\tanh\chi=0$
is caused by noise counts.}
\end{figure}

In Fig.~\ref{Fig2} we show with solid lines the dependence of the
maximal value of the Bell parameter $\mathcal{B}$ on the parameter
$\tanh\chi$ for different values of the efficiency $\eta$ and for
$N_\mathrm{nc}=10^{-6}$. For small losses this parameter exceeds the
Cirel'son bound of $2\sqrt{2}$, representing a fake violation of quantum
physics. This originates from a straightforward but inconsistent
postprocessing of the data~\cite{Lo}.

The effect of fake violations of the quantum
Bell-parameter bound does not occur when one uses
photon-number-resolving detectors, for which Eq.~(\ref{Probability})
for the probability $P_\mathrm{i_\mathrm{A},
i_\mathrm{B}}\left(\theta_\mathrm{A}, \theta_\mathrm{B}\right)$
should be rewritten as
\begin{equation}
P_\mathrm{i_\mathrm{A}, i_\mathrm{B}}\left(\theta_\mathrm{A},
\theta_\mathrm{B}\right)=\Tr\left(
\hat{\Pi}_{i_\mathrm{A}}^{(1)}\hat{\Pi}_{i_\mathrm{B}}^{(1)}
\hat{\Pi}_{j_\mathrm{A}}^{(0)}\hat{\Pi}_{j_\mathrm{B}}^{(0)}\hat\varrho\right).
\label{Probability1}
\end{equation}
In this case, Eq.~(\ref{ProbabilitySpecial}) is replaced by
\begin{eqnarray}
P_\mathrm{i_\mathrm{A}, i_\mathrm{B}}\left(\theta_\mathrm{A},
\theta_\mathrm{B}\right)&=&
\left(1-\tanh^2\chi\right)^4 e^{-4
N_\mathrm{nc}} 
\label{ProbabilitySpecialPNR}\\
&\times&\left[2\frac{C_1^2} {C_\mathrm{0}^3}
-\frac{C_\mathrm{i_\mathrm{A},i_\mathrm{B}}}{C_\mathrm{0} ^2}-
2N_\mathrm{nc}\frac{C_1}{C_\mathrm{0}^2}
+N_\mathrm{nc}^2\frac{1}{C_\mathrm{0}} \right].\nonumber
\end{eqnarray}
In addition, results obtained with photon-number-resolving detectors
demonstrate, as a rule, smaller values of the Bell parameter compared
with on-off detectors; see Fig.~\ref{Fig2} and Ref.~\cite{Semenov2} for details of
calculations.

It is worth noting that photocounting with the resolution of photon numbers
is approximately possible by splitting an initial light beam into a large number of
low-intensity beams~\cite{ArrayDetectors}. Detection of photons in each beam by an
array of photodiodes allows one to get more insight into the number statistics of the
detected photons. Similarly, one can use time multiplexing in a fiber loop with one
photodiode~\cite{LoopDetectors}. However, such measurement techniques only partly
allow one to infer the photon number statistics. The question remains whether or not
such detection  schemes are suited to completely eliminate the  fake effects under
study. This would require a more detailed analysis, which is beyond the scope of the
present paper.

\section{Double-click events}
\label{DoubleClickEvents}

In the following we will show that one can overcome the violation of the Cirel'son
bound, even when using on-off detectors. A consistent result can be obtained by the
application of a proper squash model~\cite{squashing}. The
point is that the straightforward scheme considered above discards
so-called double-click events when both detectors at the receiver
station $A(B)$ click~\cite{Lo}. 

Let us assign to the double-click events
random bits with a probability $1/2$. This changes the situation 
completely. In this case, Eq.~(\ref{Probability}) for the
probability $P_\mathrm{i_\mathrm{A},
i_\mathrm{B}}\left(\theta_\mathrm{A}, \theta_\mathrm{B}\right)$ is
replaced by
\begin{eqnarray}
P_\mathrm{i_\mathrm{A}, i_\mathrm{B}}\left(\theta_\mathrm{A},
\theta_\mathrm{B}\right)=\sum\limits_{n,m=1}^{+\infty}\Tr\left(
\hat{\Pi}_{i_\mathrm{A}}^{(n)}\hat{\Pi}_{i_\mathrm{B}}^{(m)}
\hat{\Pi}_{j_\mathrm{A}}^{(0)}\hat{\Pi}_{j_\mathrm{B}}^{(0)}
\hat\varrho\right)\nonumber\\
+\frac{1}{2}\sum\limits_{n,m,k=1}^{+\infty}\Tr\left(
\hat{\Pi}_{i_\mathrm{A}}^{(n)}\hat{\Pi}_{i_\mathrm{B}}^{(m)}
\hat{\Pi}_{j_\mathrm{A}}^{(k)}\hat{\Pi}_{j_\mathrm{B}}^{(0)}
\hat\varrho\right)\nonumber\\
+\frac{1}{2}\sum\limits_{n,m,k=1}^{+\infty}\Tr\left(
\hat{\Pi}_{i_\mathrm{A}}^{(n)}\hat{\Pi}_{i_\mathrm{B}}^{(m)}
\hat{\Pi}_{j_\mathrm{A}}^{(0)}\hat{\Pi}_{j_\mathrm{B}}^{(k)}
\hat\varrho\right)\nonumber\\
+\frac{1}{4}\sum\limits_{n,m,k,l=1}^{+\infty}\Tr\left(
\hat{\Pi}_{i_\mathrm{A}}^{(n)}\hat{\Pi}_{i_\mathrm{B}}^{(m)}
\hat{\Pi}_{j_\mathrm{A}}^{(k)}\hat{\Pi}_{j_\mathrm{B}}^{(l)}\hat\varrho\right).
\label{ProbabilitySquash}
\end{eqnarray}

\begin{figure}[ht!]
\includegraphics[clip=,width=\linewidth]{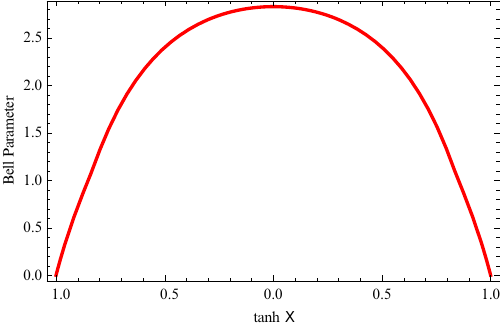}
\caption{\label{Fig3} (Color online) Maximal value of the
Bell parameter $\mathcal{B}$, obtained for lossless on-off detectors
without noise counts.}
\end{figure}

For the sake of simplicity, in the following, we only consider the
most critical 
case of lossless detectors, $\eta=1$, with no noise counts,
$N_\mathrm{nc}=0$. Under these conditions the violation of the quantum Cirel'son
bound, as considered in the previous section, attains its maximum. Substituting
state~(\ref{PDC1}) into Eqs.~(\ref{correlation}) and
(\ref{ProbabilitySquash}), one gets
\begin{equation}
E\left(\theta_\mathrm{A}, \theta_\mathrm{B}\right)=-\frac{\cos\left[
2\left(\theta_\mathrm{A}-
\theta_\mathrm{B}\right)\right]}{D},\label{correlation_squash}
\end{equation}
where
\begin{eqnarray}
D&=&1-\frac{1}{2}\tanh^2 \chi\sin^{2}\left[2\left(\theta_\mathrm{A}-
\theta_\mathrm{B}\right)\right]\label{D_on-off}\\
&+&\frac{9+3\left(1-\tanh^2\chi\right)}{2\tanh^2\chi\left(1-\tanh^2\chi\right)^2}
\nonumber\\
&\times&\left\{1-\tanh^2\chi+\frac{1}{4}\tanh^4\chi\sin^{2}\left[2\left(\theta_\mathrm
{ A}-
\theta_\mathrm{B}\right)\right]\right\}\nonumber\\
&+&\frac{\big[1-2\left(1-\tanh^2\chi\right)^2\big]\big(2-\tanh^2\chi\big)}
{\tanh^2\chi\left(1-\tanh^2\chi\right)^2}.\nonumber
\end{eqnarray}
By applying the squash model in this way, the maximum value of the Bell
parameter no longer exceeds the Cirel'son bound of $2\sqrt{2}$,
see.~Fig.~\ref{Fig3}. This result demonstrates the importance of consistent squash
models. In particular, it is necessary to include the double-click events into the
data postprocessing.

\section{Two-qubit density operator}
\label{TwoQubitDensityOperator}

In order to demonstrate the inconsistency of the
straightforward postprocessing when double-click events are
ignored, we provide the following argument. Any density
operator $\hat\rho$ of a two-qubit system, with zero mean values of
all spin projections, can be expanded into a nonorthogonal and
linearly independent basis as
\begin{equation}
\hat\rho=\frac{\hat{I}\otimes\hat{I}}{4}+\frac{1}{4}\sum\limits_{i,j=1}^{3}
E\left(\theta^{(i)}_\mathrm{A},\varphi^{(i)}_{\mathrm{A}};
\theta^{(j)}_\mathrm{B},\varphi^{(j)}_{\mathrm{B}}\right) \hspace{0.3em}
\hat{\Xi}^{(i)}_\mathrm{A}\!\otimes\hat{\Xi}^{(j)}_\mathrm{B},\label{2QB}
\end{equation}
where $\hat{I}$ is the $2\times 2$ identity matrix. $E$ denotes
the correlation coefficients for spin projections with the Euler
angles $\theta^{(i)}_\mathrm{\scriptscriptstyle
A(B)},\varphi^{(i)}_\mathrm{\scriptscriptstyle A(B)}$ $(i=1,2,3)$.
The angles $\varphi^{(i)}_\mathrm{\scriptscriptstyle A(B)}$ have
been introduced to obtain a linear-independent basis of the matrix
\begin{equation}
\hat{\Xi}^{(i)}_\mathrm{\scriptscriptstyle
A(B)}=\sum\limits_{k=1}^{3}g^{(k,i)}\left(\begin{array}{cc} \cos
2\theta^{(k)}_\mathrm{\scriptscriptstyle A(B)} &
e^{-i\varphi_\mathrm{\scriptscriptstyle A(B)}^{(k)}}
\sin 2\theta^{(k)}_\mathrm{\scriptscriptstyle A(B)}\\
e^{i\varphi_\mathrm{\scriptscriptstyle A(B)}^{(k)}} \sin
2\theta^{(k)}_\mathrm{\scriptscriptstyle A(B)}& -\cos
2\theta^{(k)}_\mathrm{\scriptscriptstyle A(B)}
\end{array}
\right),
\end{equation}
where $g^{(k,i)}$ is the metric tensor, which is inverse to
\begin{eqnarray}
\left(g^{-1}\right)_{(k,i)}&&=\cos
2\theta^{(i)}_\mathrm{\scriptscriptstyle A(B)}\cos
2\theta^{(k)}_\mathrm{\scriptscriptstyle A(B)}\\&&+ \sin
2\theta^{(i)}_\mathrm{\scriptscriptstyle A(B)}\sin
2\theta^{(k)}_\mathrm{\scriptscriptstyle
A(B)}\cos\left(\varphi_\mathrm{\scriptscriptstyle
A(B)}^{(i)}-\varphi_\mathrm{\scriptscriptstyle
A(B)}^{(k)}\right).\nonumber
\end{eqnarray}

The above-discussed mapping onto a two-qubit state
implies the identification of the coefficients
$E\left(\theta^{(i)}_\mathrm{A},\varphi^{(i)}_{\mathrm{A}};
\theta^{(j)}_\mathrm{B},\varphi^{(j)}_{\mathrm{B}}\right)$ in
Eq.~(\ref{2QB}) with relation~(\ref{correlation}), as is
usually done in experiments. To include the dependencies on the
angles $\varphi^{(i)}_\mathrm{\scriptscriptstyle A(B)}$ in the
experimental setup (see~Fig.~\ref{Fig1}), one should complete it with
phase shifters for the horizontal polarized modes. In this case one
must replace $\hat{a}_{\scriptscriptstyle \mathrm{H_{A(B)}}}$ with
$\hat{a}_{\scriptscriptstyle \mathrm{H_{A(B)}}}\exp\left(i
\varphi_{\scriptscriptstyle \mathrm{{A(B)}}}^{(i)}\right)$ in
Eqs.~(\ref{IOR1}) and (\ref{IOR2}). In the case of using on-off
detectors without considering double-click events,
Eq.~(\ref{2QB}) no longer represents a correctly defined density
operator for a two-qubit system. First, it is not uniquely defined
since it depends on the chosen angles
$\theta^{(i)}_\mathrm{\scriptscriptstyle A(B)}$ and
$\varphi_\mathrm{\scriptscriptstyle A(B)}^{(i)}$, which represent
different choices of the basis. Second and most importantly, for
some values of the angles, operator~(\ref{2QB}) appears not to be 
positive semidefinite, which contradicts the fundamental
properties of a quantum state.

\begin{figure}[ht!]
\includegraphics[clip=,width=\linewidth]{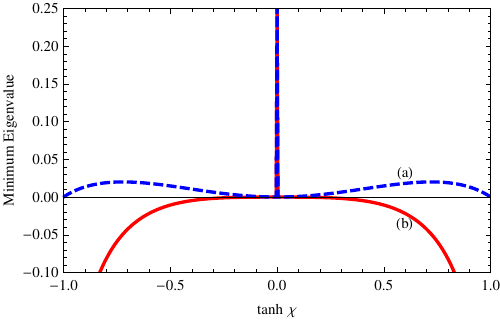}
\caption{\label{Fig4} (Color online) The minimum eigenvalue of the
density operator $\hat{\rho}$ [see~Eq.~(\ref{2QB})] vs the
parameter $\tanh \chi$ for the mean value of noise counts
$N_\mathrm{nc}=10^{-6}$, the efficiency $\eta=0.6$, and the
following sets of angles: (a)
$\theta^{(1)}_\mathrm{\scriptscriptstyle
A}=\theta^{(1)}_\mathrm{\scriptscriptstyle B}=\pi/4$,
$\varphi_\mathrm{\scriptscriptstyle
A}^{(1)}=\varphi_\mathrm{\scriptscriptstyle B}^{(1)}=0$,
$\theta^{(2)}_\mathrm{\scriptscriptstyle
A}=\theta^{(2)}_\mathrm{\scriptscriptstyle B}=\pi/4$,
$\varphi_\mathrm{\scriptscriptstyle
A}^{(2)}=\varphi_\mathrm{\scriptscriptstyle B}^{(2)}=\pi/2$,
$\theta^{(3)}_\mathrm{\scriptscriptstyle
A}=\theta^{(3)}_\mathrm{\scriptscriptstyle B}=0$,
$\varphi_\mathrm{\scriptscriptstyle
A}^{(3)}=\varphi_\mathrm{\scriptscriptstyle B}^{(3)}=0$; (b)
$\theta^{(1)}_\mathrm{\scriptscriptstyle A}=\pi/8$,
$\varphi_\mathrm{\scriptscriptstyle A}^{(1)}=0$,
$\theta^{(2)}_\mathrm{\scriptscriptstyle A}=9\pi/4$,
$\varphi_\mathrm{\scriptscriptstyle A}^{(2)}=\pi/2$,
$\theta^{(3)}_\mathrm{\scriptscriptstyle A}=\pi$,
$\varphi_\mathrm{\scriptscriptstyle A}^{(3)}=0$,
$\theta^{(1)}_\mathrm{\scriptscriptstyle B}=3\pi/15$,
$\varphi_\mathrm{\scriptscriptstyle B}^{(1)}=0$,
$\theta^{(2)}_\mathrm{\scriptscriptstyle B}=-\pi/24$,
$\varphi_\mathrm{\scriptscriptstyle B}^{(2)}=\pi/2$,
$\theta^{(3)}_\mathrm{\scriptscriptstyle B}=\pi$,
$\varphi_\mathrm{\scriptscriptstyle B}^{(3)}=0$.}
\end{figure}

In Fig.~\ref{Fig4} we show the minimum eigenvalue of the two-qubit
density operator $\hat \rho$ inferred with on-off detectors
without considering double-click events. It is clearly
seen that for different choices of the angles the minimum eigenvalue
of $\hat \rho$ can be either non-negative or it may become negative. In one
of the cases, the mapping onto a two-qubit system clearly
fails: the associated effective state does not show the properties
of a correctly defined quantum state. As a consequence, the Bell
parameter~(\ref{BellParameter}) may even exceed the quantum Cirel'son bound of
$2\sqrt{2}$. Equation~(\ref{2QB}) can be
used for reconstructing a two-qubit density operator by using the measured
correlation coefficients. If the result of this reconstruction is not positive
semidefinite,
it fails to describe a quantum state. This is an alternative possibility to
demonstrate the fake violation of quantum physics.

\section{Summary and Conclusions}
\label{SummaryAndConclusions}

Our analysis of biphoton Bell-type experiments has demonstrated that
the impossibility of discriminating between different photon numbers
is a kind of imperfection leading to surprising results. First, it
leads to a pseudoincrease of the measured Bell parameter. Second, we also
predict that for high detection efficiency the measured Bell
parameter can even show a pseudo-violation of the fundamental limit
of quantum physics, the Cirel'son bound. 

The reason for such fake violations is shown to consist in mappings of a
continuous-variable quantum state onto an
ill-defined two-qubit state that fails to be positive
semidefinite. Such fake effects disappear when one
includes in the consideration a consistent postprocessing of the
measured data by a proper squash model. The knowledge of such effects plays an
important role
in the analysis of some quantum-key-distribution protocols.

\acknowledgements

The authors gratefully acknowledge support by the Deutsche
Forschungsgemeinschaft through SFB 652. We also strongly
appreciate useful discussions with Hoi-Kwong Lo.

\end{document}